\documentclass[proof]{pasj00}

\begin{document}
\SetRunningHead{K. Enya \&  L. Abe}{A Binary Shaped Mask Coronagraph for a Segmented Pupil}
\Received{2010/07/14}

\title{A Binary Shaped Mask Coronagraph for a Segmented Pupil}

\author{Keigo \textsc{Enya}} %
\affil{Department of Infrared Astrophysics, 
  Institute of Space and Astronautical Science, \\
  Japan Aerospace Exploration Agency
  Yoshinodai 3-1-1, Sagamihara, Kanagawa 229-8510}
\email{enya@ir.isas.jaxa.jp}
\and
\author{Lyu \textsc{Abe}}
\affil{Laboratoire Hippolyte Fizeau UMR 6525 Universit\'e de Nice-Sophia 
  Antipolis Parc Valrose, F-06108 Nice, France}
\email{Lyu.Abe@unice.fr}

\KeyWords{instrumentation: high angular resolution---telescopes---stars: planetary systems } 

\maketitle

\begin{abstract}

We present the concept of a binary shaped mask coronagraph 
applicable to a telescope pupil
including obscuration, based on previous works on binary 
shaped pupil mask by \citet{Kasdin2005} and \citet{Vanderbei1999}.
Solutions with multi-barcode masks which ``skip over'' the obscuration
are shown for various types of pupil of telescope, such as SUBARU, JWST, SPICA,
and other examples.
The number of diffraction tails in the point spread function 
of the coronagraphic image is reduced to two, 
thus offering a large discovery angle.
The concept of mask rotation is also presented, 
which allows post-processing removal of diffraction tails 
and provides a 360$^{\circ}$ continuous discovery angle.
It is suggested that the presented concept offers solutions 
which potentially allow large telescopes with segmented pupil 
in future to be used as platforms for an coronagraph.

\end{abstract}

\section{Introduction}

Direct detection and spectroscopy of extra-solar planets (hereafter exo-planets) 
is expected to be one of essential methods in the understanding of the manner in which
planetary systems are born, how they evolve, and, ultimately, the identification of 
biological signatures on these planets. 
The enormous contrast in luminosity between a central star and 
an associated planet has been a critical difficulty in the direct observation 
of exo-planets. 
For instance, if the solar system is suposed to be observed from outside,
the contrast of luminosity between the Sun and Earth
is 10$^{-10}$ at visible light wavelengths and 
10$^{-6}$ in the mid-infrared wavelength region \citep{Traub2002}.
Therefore, the number of exo-planets detected directly is somewhat smaller
than those detected by other methods (\cite{Mayor1995}; \cite{Charbonneau2000}),
although the first direct observation has been finally 
achieved (\cite{Marois2008}; \cite{Kalas2008}).
The coronagraph, which was first developed for solar 
observation (\cite{Lyot1939}), is a special optics designed to reduce contrast.
It is considered that a coronagraph, designed to achieve a very high dynamic range  
has the potential to further extend the possibility of  the direct 
observation of exo-planets.

It has been considered that the performance of the
coronagraph is decreased by obscuration in the telescope pupil
which occurs from the secondary mirror, its support structure 
and the gaps between segmented mirrors.
For example, in the case of a checkerboard coronagraph (\cite{Vanderbei2004}),
which is a type of binary pupil mask coronagraph, the inner
working angle ($IWA$), with and without obscuration by
the secondary mirror, is 3$\lambda/D$ and 7$\lambda/D$ respectively, 
where $\lambda$ is the wavelength and $D$ is the diameter of the
aperture (\cite{Tanaka2006}; \cite{Enya2007}). 
In fact, off-axis telescopes are considered for 
proposed future space missions specializing in coronagraphy,
e.g., TPF-C (\cite{Traub2006}); SEE-COAST(\cite{Schneider2006}); 
PECO (\cite{Guyon2009}), in order to avoid pupil obscuration.
And therefore, various coronagraphs have been presented (e.g., summary in \cite{Guyon2006})  
for pupils without obscuration.    
If a reduction in the influence of pupil obscuration in the coronagraph
design is possible, the value of on-axis telescopes for general 
purposes (current working telescopes and those under construction) 
as platforms for a coronagraph becomes much higher. 

This paper presents  a concept of solutions to realize a coronagraph 
for a segmented pupil by employing a binary shaped pupil mask.

\section{Concept of Multi-barcode Mask Solution}

Among the various current coronagraphic methods, 
coronagraphs using binary shaped pupil masks have 
some advantages in principle.
Essentially the function of a binary pupil mask coronagraph to produce 
high contrast point spread function (PSF) is achromatic 
(except effect of scaling of PSF size),
and is somewhat less sensitive to telescope pointing error
than experienced with other coronagraphs (e.g., \cite{Lyot1939}). 
Another important property of binary pupil mask coronagraphs is the fact that 
they use only part of the pupil as the transmissive area of the mask.
In this work, we employ this property to obtain solutions
in coronagraph design with binary masks which ``skip over'' the 
obscured part of the pupil.

\begin{figure}
\vspace*{50mm}
  \begin{center} 
\FigureFile(80mm,85.4mm){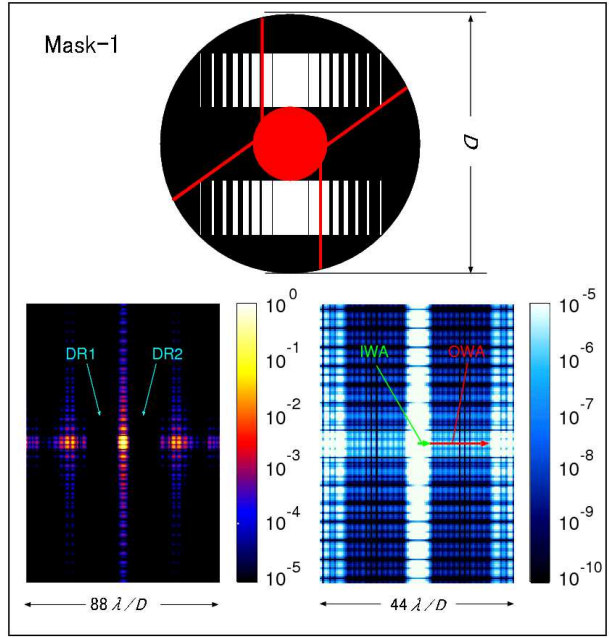}
  \end{center}
  \caption{
Top: double barcode masks (white area) applied to the pupil of a telescope
with a circular primary mirror (corresponding to black circle) 
with obscuration (red area). 
Transmissivity of the white and other areas (black and red) is 100\% and 0\%, 
respectively.
Bottom: PSF derived from simulation using the mask above.
}\label{fig1}
\end{figure}

Fig.\,\ref{fig1} shows an example of a solution for a segmented telescope pupil. 
In this case, the pupil is obscured by the
secondary mirror and four off-center support structures,
similar to the pupil of the SUBARU and GEMINI telescopes, as shown in the top 
panel of Fig.\,\ref{fig1}.
The bottom left panel shows the coronagraphic
PSF expected from the pupil mask.
Two dark regions, DR1 and DR2, are produced in the PSF. 
It must be noted that the principle of this coronagraph is essentially
the same as that of the one-dimensional coronagraph by barcode
mask presented by Kasdin et al.(2005), while the length
of the barcode mask in the vertical direction in Fig.\,\ref{fig1}
is finite, and the barcode is split into two sets (i.e., double
barcode mask), above and below the obstruction created by the secondary mirror. 
This solution has a coronagraphic power only in the horizontal direction. 
LOQO, a software presented by Vanderbei (1990), 
was used to optimize each barcode mask.
$IWA$, the outer working angle ($OWA$), and the contrast ($C_0$)
required for optimization of a one-dimensional coronagraph 
were 3.0$\lambda/D$, 16$\lambda/D$, and better than $10^{-5}$, respectively. 
In this work, the throughput  is simply defined as
the ratio of the area of transmission of the pupil, with and
without the pupil mask, which is equivalent to the ratio of
the areas of the white and black regions in the figure. 
The throughput for the solution shown in Fig.\,\ref{fig1} is 24\%. 
Values of $IWA$, $OWA$,  $C_0$, and the throughput is summarized  
in Table\,\ref{table1}, 
together with values of other solutions described below. 
The central obstruction of this solution against the coronagraph
is determined by the width of the support structure, which
is much smaller than the diameter of the obstruction due
to the secondary mirror. 
As a result, a smaller $IWA$ is realized. 
This solution has no coronagraphic power in the vertical direction. 
Along the vertical direction, the intensity of the PSF decreases, following 
the diffraction theory applied to a rectangular aperture, and there is no $OWA$.
Therefore, the contrast defined as the intensity ratio between
the core of the PSF and each position in the dark region, $C$, is not constant
($C$ is better than, or equal to $C_0$).

\section{Solutions for Various Pupils}

\subsection{Pupil Resulting from Hexagonal Mirrors}

The top of Fig.\,\ref{fig2} shows a solution using an off-centered four barcode
mask applied to a segmented telescope pupil, consisting
of hexagonal mirrors and obstructions created by a secondary
mirror and its support structure. 
This type of design of telescope is adopted by the James Webb Space Telescope (JWST). 
$IWA$,  $OWA$, and $C_0$ in this solution are 3.5$\lambda/D_{\textsl{hex}}$,  
19$\lambda/D_{\textsl{hex}}$,  
and $10^{-5.0}$, respectively,
in which $D_{\textsl{hex}}$  is defined as shown in Fig.\,\ref{fig2}. 
The throughput of the solution is 24\%. 
Off-centering the barcode mask gives rise to a peculiar shape
of the core of the PSF, while DR1 and DR2 values produced in this solution are similar 
to those when the barcode mask is not off-centered. 

JWST carries coronagraphs in two instruments, the Near-Infrared Camera 
(NIRCAM) (\cite{Rieke2005}; \cite{Green2005})
and the Mid-Infrared Instrument (MIRI) (\cite{Boccaletti2005}), 
in which Lyot-type coronagraphs and coronagraphs using 
four quadrant phase masks will be used. 
In these coronagraphs, the PSF is impaired by a complex diffraction pattern, 
especially by six bright tails in a radial direction from the core of the
PSF resulting from the segmentation of the pupil. 
As a result, the discovery angle of these coronagraphs is reduced, particularly
in positions close to the core of the PSF. 
These coronagraphs use devices set at the focal plane 
in order to realize a high contrast image 
(i.e., at the occluting mask or four quadrant phase mask), 
so that these coronagraphs are, in principle, sensitive to telescope
pointing error and have limited working bandwidth. 
If it is possible to use a binary pupil mask coronagraph, 
these limits are essentially relaxed and the discovery angle
of the coronagraphic image  can be improved. 
On the other hand, a solution using a binary pupil mask, as shown 
in Fig.\,\ref{fig2},  applies a constraint of $OWA$.
In order to get the best coronagraph design for each mission, 
it is essential to estimate the expected observational
performance from both the instrument specification
and scientific simulation.

\subsection{Pupil with Central Obscuration and On-axis Spiders}

The bottom of Fig.\,\ref{fig2} shows a solution provided by a double barcode
mask, applied to an on-axis  telescope pupil with obscuration by the secondary
mirror and its four on-axis support structures.
This is an example of a solution obtained from optimization presuming a 
central obstruction of the barcode mask.
It should be noted that the central obstruction in this case is
caused by the width of the support structure (not by the diameter of the
secondary mirror). 
$IWA$,  $OWA$, and $C_0$ in this solution are 3.4$\lambda/D_{\textsl{hex}}$,  
15$\lambda/D_{\textsl{hex}}$,  and $10^{-5.0}$, respectively. 
The throughput of the solution is 15\%. 

$C$ and $IWA$ of this solution satisfy
the requirement for a mid-infrared coronagraph  (\cite{Enya2010}) for the
Space Infrared telescope for Cosmology and Astrophysics (SPICA) (\cite{Nakagawa2009}). 
SPICA will carry an on-axis Ritchey-Chretien telescope with a 3m class diameter aperture, 
and it is planned to be launched in 2018. 
Use of a binary-shaped pupil mask is considered as the baseline 
solution for the SPICA coronagraph because of its achromatic work, 
robustness against telescope pointing error caused by vibration of 
cryo-coolers and other mechanics, and feasibility.

\subsection{Further Variations}

Fig.\,\ref{fig4} shows further variations of solutions consisting of
multi barcode masks.
Mask-4 is a solution for a telescope pupil which is the same as the pupil
in the case of Mask-2.
Four barcode masks used in Mask-4 and Mask-2 are common.
In addition, four segments of the pupil, located to the left and right 
of the central obscuration, are used in order to demonstrate the improvement
in the throughput.
$IWA$,  $OWA$, and $C_0$ in this solution are 3.9$\lambda/D_{\textsl{hex}}$, 
14$\lambda/D_{\textsl{hex}}$,  and $10^{-5.0}$, respectively. 
The throughput of this solution is 27\%. 
Optimization of these newly used segments was carried out
with the constraint of central obscuration, as in the case of Mask-3.
In comparison with Mask-2, this solution can be regarded as an example  
in which $C_0$ is maintained but $IWA$ and $OWA$ is compromised  
as the result of trade-off.

Mask-5 is a solution for a telescope pupil which is same as the pupil
in the case of Mask-3.
Eight barcode masks were employed in order to extend the effective area 
used by masks. 
Optimizations of each mask were carried out 
with constraints in the central obscuration caused by
the secondary mirror or its support structure.
$IWA$,  $OWA$, and $C_0$ in this solution are 3.3$\lambda/D$, 10$\lambda$,  
and $10^{-5.3}$, respectively. 
The throughput of this solution is 30\%, implying a large improvement
over the case of Mask-3. 
In comparison with Mask-3, this solution can be regarded as an example  
in which $IWA$ is maintained but $C_0$ and $OWA$ is compromised  
as the result of trade-off.
 
Further improvement in performance is possible,
if the telescope design takes account of use of a multi-barcode 
pupil mask coronagraph.  
Mask-6 and Mask-7 are solutions presuming segmented rectangular mirrors.
The throughput of these solutions are 50\%, which is the highest 
of the solutions presented in this paper. 
For Mask-6, $IWA$,  $OWA$, and $C_0$ in this solution are 3.6$\lambda/D_{\textsl{rect}}$, 
11$\lambda/D_{\textsl{rect}}$,  and $10^{-6.0}$, respectively,
in which $D_{\textsl{rect}}$ is defined as shown in Fig.\,\ref{fig3}. 
Mask-7 provides $IWA$=2.5$\lambda/D_{\textsl{rect}}$, which is significantly 
better than Mask-6, while $C_0$ and $OWA$ are common to
Mask-6 and Mask-7.

\section{Discussion and Summary}

\begin{figure}
  \begin{center}
\vspace*{50mm}
    \FigureFile(80mm, 106.9mm){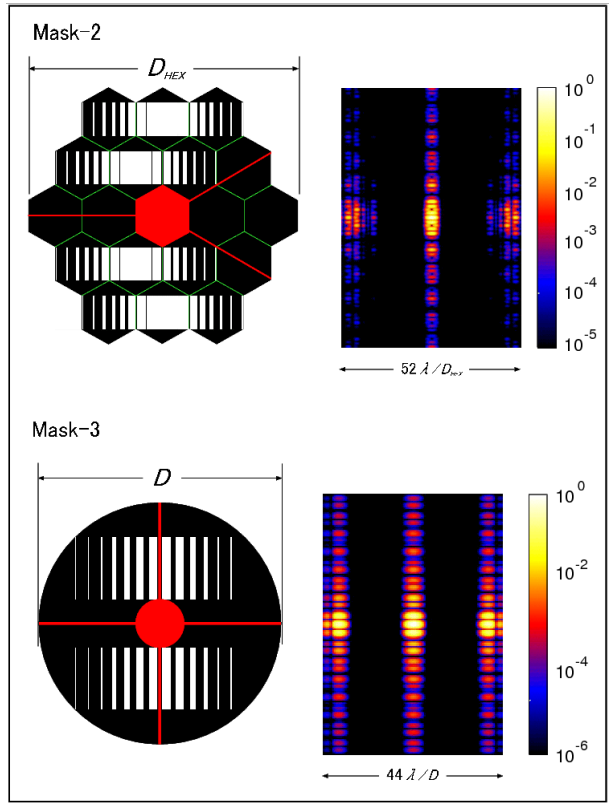}   
  \end{center}
  \caption{
Multi barcode masks and PSF derived from simulation.
The green line in the Mask-2 shows the join between segments.
}\label{fig2}
\end{figure}

\begin{figure}
  \begin{center}
\vspace*{50mm}
     \FigureFile(160mm, 107.7mm){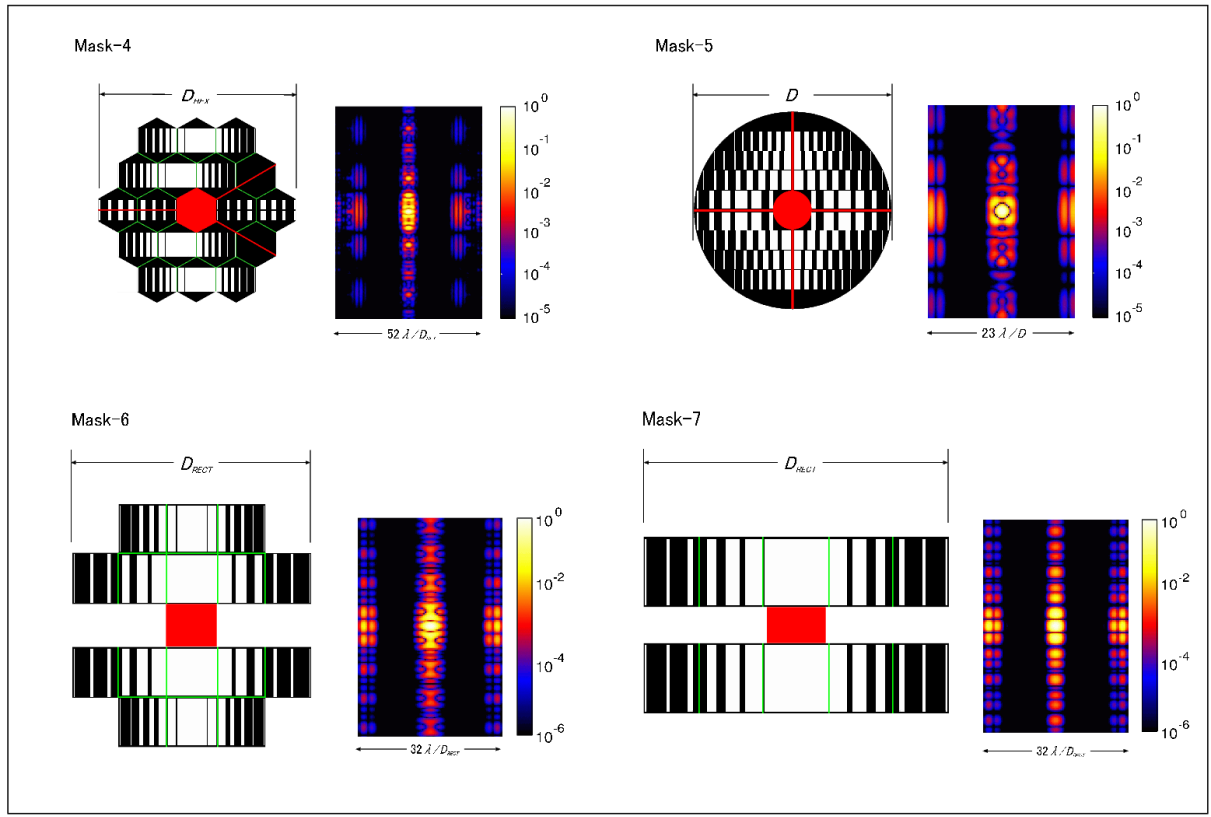}
  \end{center}
  \caption{
Multi barcode masks and PSF derived from simulation.
}\label{fig3}
\end{figure}

\begin{figure}
  \begin{center}
\vspace*{50mm}
 \FigureFile(80mm, 101.6mm){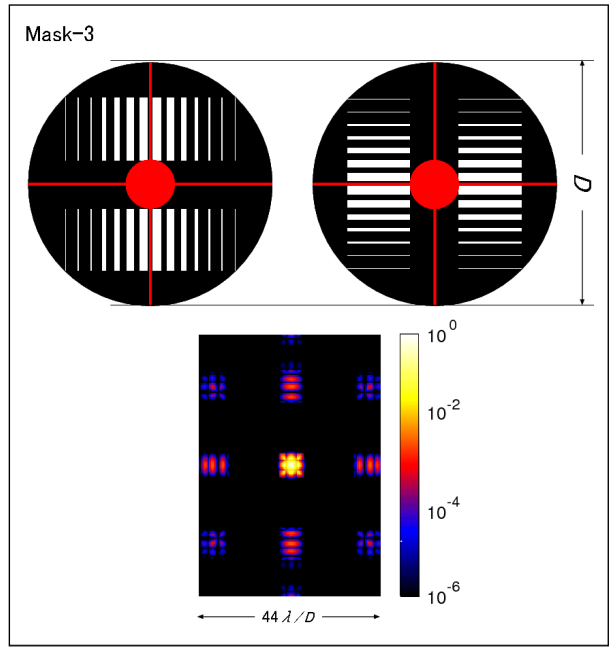}
  \end{center}
  \caption{
Concept of mask rotation. 
Top: mask configuration, before and after the rotation.
Bottom: composite PSF, having a dark region as the sum
of each of the dark regions of PSF, before and after the mask rotation.
}\label{fig4}
\end{figure}

If rotation of the pupil mask and coronagraphic imaging, before and after 
the rotation, are possible, the total discovery angle is improved. 
Fig.\,\ref{fig4} shows the concept 
of rotating the pupil mask by 90 degrees, applied to the
solution shown in Fig.\,\ref{fig2}. 
As a result of double coronagraphic imaging, before and after mask rotation, 
most of the influence of the diffraction tails in the coronagraphic PSF is removed,
and totally 360$^{\circ}$ continuous discovery angle is provided. 
The improved discovery angle makes it possible to observe companions 
of a central star, even if the companions are buried in diffraction 
tails of the original PSF. 
The mask rotation technique can be especially useful for SPICA, 
in which the role of the telescope (\cite{Trauger2007})  
is strongly constrained due to 
the thermal system design  required to realize a cryogenic 
infrared telescope satellite utilizing radiation cooling.

A binary shaped pupil mask has also been used in order to 
support a Phase Induced Amplitude Apodization (PIAA)
coronagraph (\cite{Guyon2010}).
It is, in principle, possible with such a hybrid coronagraph
to realize smaller $IWA$ and higher throughput than 
the values in coronagraph employing only a binary shaped pupil mask.
The currently presented hybrid coronagraph by \citet{Guyon2010}
includes circular apodization produced by PIAA and a binary shaped mask 
consisting of concentric rings (\cite{Vanderbei2003}).
Therefore, the coronagraphic power is along radial direction in PSF 
and is not one-dimensional.
We would like to point out the potential
to combine barcode masks and a one-dimensional PIAA
in order to realize a one-dimensional hybrid coronagraph
which is applicable to segmented telescope pupils
like those shown in this paper.

In general, the size of a telescope aperture is limited by various factors.
For instance, the size of the rocket fairing constrains  
off-axis space telescopes with seamless mirrors proposed especially 
for the observation of exo-planets.
If a segmented pupil becomes more useful for coronagraphic observation
of exo-planets, on-axis telescopes with a segmented pupil for 
general purposes becomes more valuable as a coronagraph platform.
In the case of ground based telescopes, the current largest class of telescopes
(e.g. VLT, KECK, GEMINI, SUBARU and so on), 
can be good target for application of a one-dimensional 
coronagraph using a binary shaped mask.
These telescopes are starting direct detections of giant, young
planetary objects in near infrared
(\cite{Chauvin2004}; \cite{Marois2008}; \cite{Lagrange2010}; \cite{Thalmann2010}).
Giant ground based telescope of the futre (e.g., TMT, EELT)
will extend these observations in spatial resolution and sensitivity.

Space telescopes have further potential, especially for observation
in mid-infrared wavelength region.  
The contrast provided by several of solutions presented in this paper is 
$\sim10^{-6}$, which is the contrast needed for observation of 
matured terrestrial
exo-planets in the mid-infrared (\cite{Traub2002}). 
This fact suggests that a mid-infrared coronagraph with
giant telescope, consisting of segmented mirrors, has the potential
for terrestrial planet search in future,
where there is a critical spectral feature, O$_3$\,(9.8$\mu$m)
considered to be a biomarker in atmosphere of terrestrial planets.
In contrast, coronagraphic search for terrestrial planets 
in the visible wavelength requires observation with much higher 
contrast, $10^{-10}$  (\cite{Traub2002}), which any of the
masks in this paper are unable to reach.
Proposed off-axis telescopes with a seamless pupil, 
e.g., TPF-C (\cite{Traub2006}); SEE-COAST (\cite{Schneider2006}); 
PECO (\cite{Guyon2009}) 
might be a reasonable solution to attain such ultimate 
contrast, rather than larger telescopes with a segmented pupil.

This paper presents a one-dimensional coronagraphic solution for a
segmented telescope pupil by applying a type of binary shaped mask,
a multiple barcode masks,
which ``skip over'' the obscured part of the pupil.   
These coronagraphs have the general advantage of binary pupil mask coronagraphs,
i.e., lack of susceptibility to telescope pointing errors and
less constraint on the bandwidth. 
Furthermore, the multi-barcode mask coronagraph provides 
a large discovery angle and a small $IWA$, even for a pupil with a
large central obstruction. 
We suggest  that the concept of these solutions has a potential use
in facilitating large telescopes having a segmented pupil 
to be used as platforms for an advanced coronagraph.


\bigskip

We deeply thank to pioneers of the barcode mask, 
particularly N. J. Kasdin and R. J. Vanderbei, with the best respect.
The work is  supported by the Japan Society for 
the Promotion of Science, and the Ministry of Education, Culture, 
Sports, Science and Technology of Japan. 
We would like to express special gratitude to S. Tanaka,
even after the change of his field.



\newpage

\begin{table}
\vspace*{70mm}
  \label{table1}
  \begin{center}
\caption{Parameters in mask designs.}
     \begin{tabular}{lcllc}
      \hline 
      \,\,\,Mask   & $C_0$    & \,\,\,\,\,$IWA$ & \,\,\,$OWA$ & Throughput (\%) \\
      \hline 
      Mask-1    & $10^{-5.0}$   & 3.0\,$\lambda/D$                  & 16\,$\lambda/D$        & 24  \\
      Mask-2    & $10^{-5.0}$   & 3.5\,$\lambda/D_{\textsl{hex}}$   & 19\,$\lambda/D_{\textsl{hex}}$  & 24  \\
      Mask-3    & $10^{-6.0}$   & 3.4\,$\lambda/D$                  & 15\,$\lambda/D$        & 15  \\
      Mask-4    & $10^{-5.0}$   & 3.9\,$\lambda/D_{\textsl{hex}}$   & 14\,$\lambda/D_{\textsl{hex}}$  & 27  \\
      Mask-5    & $10^{-5.3}$   & 3.3\,$\lambda/D$                  & 10\,$\lambda/D$        & 30  \\
      Mask-6    & $10^{-6.0}$   & 3.6\,$\lambda/D_{\textsl{rect}}$  & 11\,$\lambda/D_{\textsl{rect}}$ & 50  \\
      Mask-7    & $10^{-6.0}$   & 2.5\,$\lambda/D_{\textsl{rect}}$  & 11\,$\lambda/D_{\textsl{rect}}$ & 50  \\
      \hline 
      \end{tabular}
  \end{center}
\end{table}


\clearpage

\vspace*{50mm}

\clearpage

\clearpage

\clearpage

\end{document}